\documentclass[journal]{IEEEtran}

\usepackage[T1]{fontenc}

\usepackage{siunitx}

\usepackage{tikz}
\usepackage{pgfplots}
\pgfplotsset{compat=1.18}
\usetikzlibrary{calc}
\usetikzlibrary{fadings}
\usetikzlibrary{positioning}
\usetikzlibrary{svg.path}
\usepackage{cite}
\usepackage{orcidlink}

\begin{document}
\bstctlcite{IEEEexample:BSTcontrol}
\title{A Structural Analysis of the User Behavior Dynamics for Environmentally Sustainable ICT}

\author{Stefan Roth\,\orcidlink{0000-0003-4408-3306} and Aydin Sezgin\,\orcidlink{0000-0003-3511-2662}%
\thanks{This work was supported in part by the German Federal Ministry of Education and Research (BMBF) project 6G-ANNA [grant agreement number 16KISK095], and in part by the German Federal Ministry of Education and Research (BMBF) in the course of the 6GEM Research Hub under grant 16KISK037.}%
\thanks{S. Roth and A. Sezgin are with the Institute of Digital Communication Systems, Ruhr University Bochum, 44801 Bochum, Germany (e-mail: stefan.roth-k21@rub.de, aydin.sezgin@rub.de).}}

\maketitle

\begin{abstract}
The sector of information and communication technology (ICT) can contribute to the fulfillment of the Paris agreement and the sustainable development goals (SDGs) through the introduction of sustainability strategies. For environmental sustainability, such strategies should contain efficiency, sufficiency, and consistency measures. To propose such, a structural analysis of ICT is undertaken in this manuscript. Thereby, key mechanisms and dynamics behind the usage of ICT and the corresponding energy and resource use are analyzed by describing ICT as a complex system. The system contains data centers, communication networks, smartphone hardware, apps, and the behavior of the users as sub-systems, between which various Morinian interactions are present. Energy and non-energy resources can be seen as inputs of the system, while e-waste is an output. Based on the system description, we propose multiple measures for efficiency, sufficiency and consistency to reduce greenhouse gas emissions and other environmental impacts. 
\end{abstract}


\IEEEpeerreviewmaketitle

\section{Introduction}

\begin{figure}
    \centering
    \input{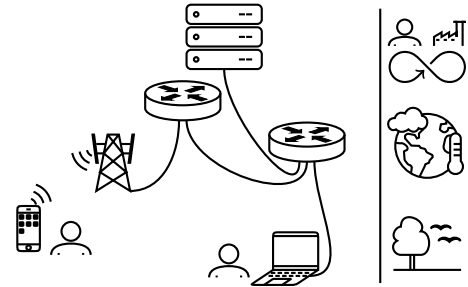}
    \caption{ICT consists of user devices, communication networks, and data centers with the according hardware and software components. Within this work, the aim is to design measures for environmentally sustainable ICT to achieve sustainable production and consumption, climate action and protection of life on land.}
    \label{fig:sdg-illustration}
\end{figure}

\IEEEPARstart{I}{nformation} and communication technology (ICT) is estimated to be responsible for 1.4\%-3.9\% of the current greenhouse gas emissions \cite{MALMODIN2024102701,BELKHIR2018448,FREITAG2021100340}. While the global climate targets require decarbonization \cite{IEA2023}, multiple studies estimate that the energy usage of ICT continues increasing \cite{FREITAG2021100340,MALMODIN2024102701,BELKHIR2018448,Andrae2020}, and some studies also predict increasing greenhouse gas emissions \cite{FREITAG2021100340,BELKHIR2018448}. In addition to the greenhouse gas emissions, also the mining of resources included within the ICT devices and e-waste have impacts on ecosystems and humans \cite{FREITAG2021100340,https://doi.org/10.1002/lno.11403,GlobalEWasteMonitor2024}. To fulfill the Paris agreement \cite{paris2015} and the sustainability development goals (SDGs) \cite{Agenda2030}, our societies need to undergo various changes, also affecting ICT. According to the Paris agreement, all countries should implement policies reducing greenhouse gas emissions to limit global heating at well-below $\SI{2}{\degree C}$, while targeting to limit global heating at $\SI{1.5}{\degree C}$ and ensuring a just transition. Within the SDGs, the carbon emissions and resource usage of ICT are covered within the environmental targets on climate action (goal 13) and life on land (goal 15). Additionally, sustainable patterns for consumption and production should be implemented (goal 12). However, the fulfillment of these SDGs is in wide parts off-track \cite{United_Nations_Department_of_Economic_and_Social_Affairs2023-vz} and six out of nine planetary boundaries are transgressed \cite{doi:10.1126/sciadv.adh2458}. To still achieve the above goals or at least minimize the overshoot, significantly increased effort is needed \cite{sru2024}. This is also core of the proposed industry 5.0, which is currently discussed at the EU level to re-adjust the industry targets to be aligned with the SDGs \cite{doi/10.2777/17322}. If ICT should contribute with a fair share to the required efforts, reductions of the greenhouse gas emissions of different sub-sectors between 37\% and 67\% should be achieved between 2020 and 2030 \cite{ItuScienceBasedTargets}, and according to the latest data even higher reductions might be required \cite{doi/10.2777/17322}. In this work, we aim at proposing such environmental sustainability strategies, as illustrated in \figurename~\ref{fig:sdg-illustration}.

\begin{figure}
    \centering
    \input{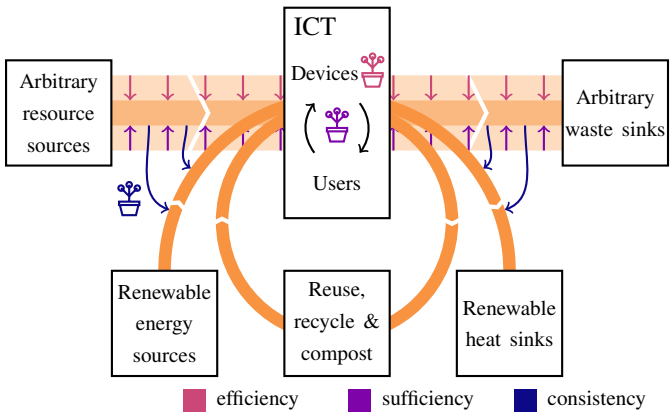}
    \caption{Flows of the resources from their sources to waste sinks. Efficiency measures target to change technical components to reduce the resource usage, while sufficiency measures target to achieve changes within the demand to limit the resource use. Consistency measures additionally aim to close material loops and change the energy usage to renewable sources.} 
    \label{fig:consistency_efficiency_sufficiency_illustration}
\end{figure}

The environmental implications of ICT are related to data centers, communication networks and user devices such as smartphones \cite{BELKHIR2018448}. To fulfill the environmental targets, different sustainability measures are needed for ICT that jointly fulfill the requirements. Therefore, a significant emphasis in the research on ICT has been laid on increasing the \emph{efficiency} of various technical components of the ICT devices and their communication \cite{Santarius2023}. Within efficiency measures, each ICT component is optimized to minimize the energy or resource consumption under the assumption of a constant usage.  Within electronic circuits, a smaller transistor size has lead to a more energy-efficient computing. For the device's communication, enhancements in beamforming and hardware design have lead to significant energy efficiency enhancements in 5G mobile networks \cite{EricssonEnergyCurve}. However, also efficient systems can have a high resource consumption, such as through usage patterns related to a high energy use. For example, technological developments have lead to changes in user behavior, which impacts the energy usage \cite{Bates2015/09}. Hence, \emph{sufficiency} measures are needed alongside with the efficiency measures \cite{Santarius2023,ItuScienceBasedTargets}. Such measures aim to reduce the overall resource consumption through avoiding a resource-intensive demand while providing everyone with a sufficient level of resources. To design sufficient systems, it is required to take the dynamics behind the user behavior and its changes into account. Recently, \cite{sru2024} has highlighted the general necessity of sufficiency measures and the criticality of implementing such in a socially just way. In addition to efficiency and sufficiency measures, the flows of the resources from their sources to their sinks should be considered in the design of sustainability strategies. This is done by \emph{consistency} measures, which aim to achieve environmental friendly systems through closing material loops and using renewable energy. For example, such measures can target a reduced usage intensity, re-usage of devices or hardware components, or increased device recycling rates. Therefore, multiple companies aim to change the energy to renewable sources that ICT devices consume during their usage. Moreover, various approaches exist to collect old devices such that these become available for recycling \cite{EuropeanEconomicAndSocialCommittee.CEPS.2019,BAI2018227}. The mechanisms of all three kinds of measures for environmental sustainability are illustrated in \figurename~\ref{fig:consistency_efficiency_sufficiency_illustration}.

To propose sustainability measures for ICT that cover sufficiency and consistency, systems thinking is required \cite{zora232410,doi/10.2777/17322}. This allows taking into account the structural effects of the social context in which the technology is applied and the interaction of different ICT components with each other. Further, the resource sources and waste sinks can be taken into account. Note that the ecological implications of each individual ICT device are mainly determined by its lifetime, the energy and resources required for production, the use phase energy consumed by the device, and the end-of-life treatment \cite{BELKHIR2018448}. The largest portion of the growth of the energy demand of ICT can be allocated to communication networks and data centers \cite{FREITAG2021100340,Andrae2020}. Additionally, according to \cite{BELKHIR2018448}, also smartphones may have increasing greenhouse gas emissions. For user devices such as smartphones and laptops, the largest portion of the environmental impact comes from embodied emissions, i.e., emissions coming from the devices' production \cite{FREITAG2021100340}. Thereby, communication networks and data centers are only used indirectly by the users, i.e., users access apps or websites on their smartphones or laptops, which then connect through communication networks to data centers. Hence, we describe the different ICT components together as a complex system \cite{Fieguth2021}, aiming to cover the most relevant structural aspects for environmental sustainability. Based on this analysis, we propose different measures targeting environmental sustainability. 

\subsection{Related Works}

The works \cite{BELKHIR2018448,su10093027,MALMODIN2024102701} have estimated the greenhouse gas emissions and electricity consumption of ICT based on a life-cycle assessment. Similarly, the works \cite{challe6010117,Andrae2020} have estimated the electricity consumption of ICT by estimating the electricity consumption within the use phase and applying a life cycle ratio to consider the production phase within the estimates. In \cite{FREITAG2021100340}, the assumptions behind previous estimates of the greenhouse gas emissions of ICT from \cite{challe6010117,BELKHIR2018448,su10093027} and other studies from the same research groups have been analyzed, and adjusted estimates have been suggested. The international energy agency (IEA), which tracks the carbon emissions and electricity usage for a wide number of sectors, uses the estimates from \cite{MALMODIN2024102701,su10093027} within their analysis \cite{IEADataCentersTransmissionNetworks}. Hence, the IEA reports data centers and communication networks to be responsible for 1\% of all greenhouse gas emissions \cite{IEADataCentersTransmissionNetworks}. Moreover, \cite{IEADataCentersTransmissionNetworks} describes that more efforts are needed to align data centers and communication networks with the IEA's Net Zero Emissions by 2050 scenario, and suggests some measures to achieve this.

To reduce the greenhouse gas emissions in line with the global goals, \cite{ItuScienceBasedTargets} found that reductions of the different sub-sectors of ICT between 37\% and 67\% are needed between 2020 and 2030. Moreover, \cite{ItuScienceBasedTargets} suggested that efficiency, sufficiency, and consistency measures are needed jointly to achieve these reduction targets. Accordingly, a high number of studies has proposed measures for a sustainable ICT or ICT for sustainability. Thereby, the term sustainable ICT refers to a sustainable design of ICT itself, while ICT for sustainability refers to the usage of ICT in other sectors to enable sustainable practices. Several of these studies have been analyzed in \cite{Santarius2023}. Thereby, it was found that various studies exist on energy efficiency measures for ICT and enablers, while multiple blind spots were found in the areas of sufficiency and consistency measures and structural effects.

The rebound effect describes a situation where resource savings due to efficiency measures lead to behavior changes, which result in more resources being used. In \cite{10.1007/978-3-319-09228-7_26}, different rebound effects induced by ICT were analyzed and classified as direct rebound effects, indirect rebound effects and economy-wide rebound effects. Among such, the impact of the introduction of technological devices into households on behavioral changes was investigated in \cite{Bates2015/09}. Thereby, it was found that the availability of more advanced technology can lead to an extended use of networked devices. Similarly, in \cite{WILLIAMS2022112033}, various possible energy use implications of 5G were analyzed. Thereby, it was found that the deployment of new cellular networks is at least partially responsible for increases in demand. To address these rebound effects, \cite{zora232410} suggests to employ systems thinking.

The basic principles of the thinking in systems and complex systems are described in \cite{Meadows2008Thinking,Fieguth2021}. In \cite{Park2005TheIA}, the internet is modeled as a complex system, focusing on network topology and routing protocols. Based on this, various errors potentially occurring during data transmissions were identified. In \cite{8259613}, the development of individual software projects is described as a complex system, involving a high number of different people with different roles that are organized in teams. In a broader context, a complex system description has also been used to identify limits to growth \cite{Meadows1972} and various aspects of the organization of cities and transportation \cite{MARCHETTI199475}.

\subsection{Contribution}

In this work, we contribute to addressing the blind spots in the research described in \cite{Santarius2023}. Therefore, as suggested by \cite{zora232410,doi/10.2777/17322}, we analyze the structural effects of the ICT usage by applying systems thinking and describing ICT as a complex system. Furthermore, we use the system description to propose various sustainability measures that cover all three categories. The main contributions of this study are as follows:
\begin{itemize}
    \item We describe ICT as a complex system covering the interaction of the users with the technology and various system inputs and outputs. Therefore, we critically review the academic literature, various industry reports and data sources. Based on this, we identify the key mechanisms of ICT that are responsible for user behavior changes and the consumption of energy and other resources. 
    \item Based on the aforementioned system description, various points are identified where sustainability measures can be implemented that take structural aspects into account. To reduce the greenhouse gas emissions and other environmental impacts, we propose multiple measures targeting efficiency, sufficiency and consistency. Thereby, the use phase and the production phase are both taken into account. 
\end{itemize}

\section{Describing ICT as a Complex System}\label{sec:IctAsComplexSystem}

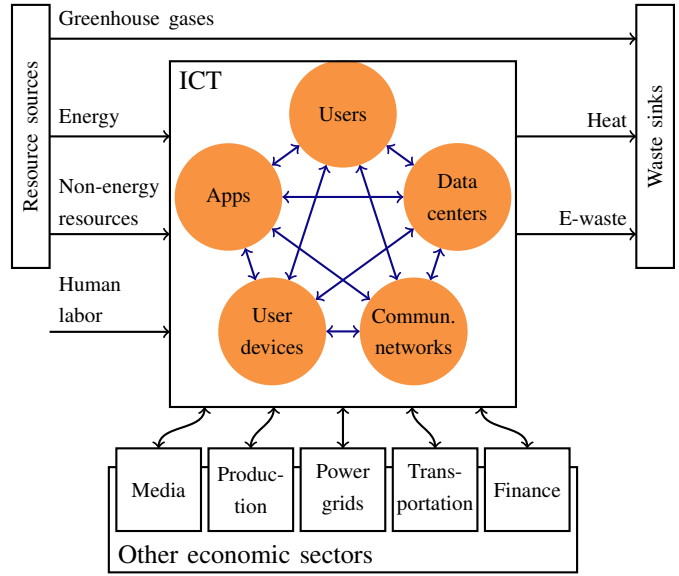
\begin{figure}
    \centering
    \definecolor{mycolor1}{rgb}{0.97223,0.58172,0.25408}%
\definecolor{mycolor2}{rgb}{0.79638,0.27798,0.47132}%
\definecolor{mycolor3}{rgb}{0.49343,0.01152,0.65803}%
\definecolor{mycolor4}{rgb}{0.05038,0.02980,0.52797}%

\begin{tikzpicture}
    
    \node[text width=3.5cm,minimum height=0.5cm,align=center,draw,thick,inner sep=0,outer sep=0,fill=white,rotate=90] (r_in) at (-4.15,1.3) {\footnotesize Resource sources};
    \node[text width=3.5cm,minimum height=0.5cm,align=center,draw,thick,inner sep=0,outer sep=0,fill=white,rotate=90] (r_out) at (4.15,1.3) {\footnotesize Waste sinks};

    \node (in0) at (-3.9,2.6) {};
    \node (in1) at (-3.9,1.3) {};
    \node (in2) at (-3.9,0) {};
    \node (in3) at (-3.9,-1.3) {};
    %\node (in4) at (-3.9,-1.9) {};

    \draw[thick, ->] (in1.center) -- ++(1.6,0); 
    \draw[thick, ->] (in2.center) -- ++(1.6,0); 
    \draw[thick, ->] (in3.center) -- ++(1.6,0);
    %\draw[thick, ->] (in4.center) -- ++(1.6,0);
    
    \node[anchor=south west,text width=1.7cm] at (in1) {\footnotesize Energy}; 
    \node[anchor=south west,text width=1.7cm] at (in2) {\footnotesize Non-energy\\resources}; 
    \node[anchor=south west,text width=1.7cm] at (in3) {\footnotesize Human\\labor}; 
    %\node[anchor=south west,text width=1.7cm] at (in4) {\footnotesize Capital\\investment}; 

    \node (out0) at (3.9,2.6) {};
    \node (out1) at (3.9,1.3) {};
    \node (out2) at (3.9,0) {};
    %\node (out3) at (3.9,-1.9) {};
    
    \draw[thick, <-] (out1.center) -- ++(-1.6,0); 
    \draw[thick, <-] (out2.center) -- ++(-1.6,0); 
    %\draw[thick, <-] (out3.center) -- ++(-1.6,0); 
    
    \node[anchor=south east,text width=1.7cm,align=right] at (out1) {\footnotesize Heat}; 
    \node[anchor=south east,text width=1.7cm,align=right] at (out2) {\footnotesize E-waste};
    %\node[anchor=south east,text width=1.7cm,align=right] at (out3) {\footnotesize Dividend\\payout}; 

    \draw[thick, ->] (in0.center) -- (out0.center); 
    
    \node[anchor=south west,text width=6cm] at (in0) {\footnotesize Greenhouse gases}; 
    
    \node[text width=1.3cm,minimum height=1.43cm,align=center,fill=mycolor1,inner sep=0,outer sep=0,circle] (c1) at (378:1.6cm) {\footnotesize Data\\centers};
    \node[text width=1.3cm,minimum height=1.43cm,align=center,fill=mycolor1,inner sep=0,outer sep=0,circle] (c2) at (306:1.6cm) {\footnotesize Commun.\\networks};
    \node[text width=1.3cm,minimum height=1.43cm,align=center,fill=mycolor1,inner sep=0,outer sep=0,circle] (c3) at (234:1.6cm) {\footnotesize User\\devices};
    \node[text width=1.3cm,minimum height=1.43cm,align=center,fill=mycolor1,inner sep=0,outer sep=0,circle] (c4) at (162:1.6cm) {\footnotesize Apps};
    \node[text width=1.3cm,minimum height=1.43cm,align=center,fill=mycolor1,inner sep=0,outer sep=0,circle] (c5) at (90:1.6cm) {\footnotesize Users};

    \draw[<->,thick,mycolor4] (c1)--(c2);
    \draw[<->,thick,mycolor4] (c1)--(c3);
    \draw[<->,thick,mycolor4] (c1)--(c4);
    \draw[<->,thick,mycolor4] (c1)--(c5);
    \draw[<->,thick,mycolor4] (c2)--(c3);
    \draw[<->,thick,mycolor4] (c2)--(c4);
    \draw[<->,thick,mycolor4] (c2)--(c5);
    \draw[<->,thick,mycolor4] (c3)--(c4);
    \draw[<->,thick,mycolor4] (c3)--(c5);
    \draw[<->,thick,mycolor4] (c4)--(c5);

    \draw[thick] (-2.3,-2.3) rectangle (2.3,2.3);
    \node[anchor=north west] at (-2.3,2.3) {ICT};

    %\node[text width=1.3cm,minimum height=1.43cm,align=center,draw,dashed,inner sep=0,outer sep=0,circle] (c1) at (378:1.65cm) {};
    %\node[text width=1.3cm,minimum height=1.43cm,align=center,draw,dashed,inner sep=0,outer sep=0,circle] (c2) at (306:1.65cm) {};
    %\node[text width=1.3cm,minimum height=1.43cm,align=center,draw,dashed,inner sep=0,outer sep=0,circle] (c3) at (234:1.65cm) {};
    %\node[text width=1.3cm,minimum height=1.43cm,align=center,draw,dashed,inner sep=0,outer sep=0,circle] (c4) at (162:1.65cm) {};
    %\node[text width=1.3cm,minimum height=1.43cm,align=center,draw,dashed,inner sep=0,outer sep=0,circle] (c5) at (90:1.65cm) {};

    \draw[thick] (-3.1125,-3.1) rectangle (3.1125,-4.5);
    \node[anchor=south west] at (-3.1125,-4.5) {Other economic sectors};

    \node[text width=1.125cm,minimum height=1.1cm,align=center,draw,thick,inner sep=0,outer sep=0,fill=white] (s1) at (-2.45,-3.4) {\footnotesize Media};
    \node[text width=1.125cm,minimum height=1.1cm,align=center,draw,thick,inner sep=0,outer sep=0,fill=white] (s2) at (-1.225,-3.4) {\footnotesize Produc-tion};
    \node[text width=1.125cm,minimum height=1.1cm,align=center,draw,thick,inner sep=0,outer sep=0,fill=white] (s3) at (0,-3.4) {\footnotesize Power\\grids};
    \node[text width=1.125cm,minimum height=1.1cm,align=center,draw,thick,inner sep=0,outer sep=0,fill=white] (s4) at (1.225,-3.4) {\footnotesize Trans-portation};
    \node[text width=1.125cm,minimum height=1.1cm,align=center,draw,thick,inner sep=0,outer sep=0,fill=white] (s5) at (2.45,-3.4) {\footnotesize Finance};

    \draw[<->,thick] (s1) to [out=90,in=-90] ($(-2.3,-2.3)!0.1!(2.3,-2.3)$);
    \draw[<->,thick] (s2) to [out=90,in=-90] ($(-2.3,-2.3)!0.3!(2.3,-2.3)$);
    \draw[<->,thick] (s3) to [out=90,in=-90] ($(-2.3,-2.3)!0.5!(2.3,-2.3)$);
    \draw[<->,thick] (s4) to [out=90,in=-90] ($(-2.3,-2.3)!0.7!(2.3,-2.3)$);
    \draw[<->,thick] (s5) to [out=90,in=-90] ($(-2.3,-2.3)!0.9!(2.3,-2.3)$);
    
\end{tikzpicture}
    \caption{ICT can be modeled as a complex system.}
    \label{fig:complex_system_model_drawing}
\end{figure}

From a systemic perspective, smartphone hardware, apps, communication networks, data centers, and the behavior of the users can be seen as interconnected systems. Thereby, even though no central coordination of smartphone developers, app designers, network engineers, data center providers, and users exists, couplings between almost all of the sub-systems are present. We will show later that, due to the specific kind of couplings, ICT is a complex system \cite{Estrada2023}, i.e., a system of systems \cite{Fieguth2021} that has all of the aforementioned systems as sub-systems (see \figurename~\ref{fig:complex_system_model_drawing}). Capital investments enable the usage of energy, non-energy resources and human labor to build up, extend, and maintain the ICT hardware, software, production facilities for the devices, etc. During the usage of the devices, additional energy is used. Hence, the relevant inputs of the system are energy, non-energy resources and human labor, which are all required within the development of ICT. Thereby, the energy use and extraction of non-energy resources are both related to the emission of greenhouse gases. Due to devices becoming depreciated or discarded \cite{Meadows1972}, e-waste leaves the system. Furthermore, the energy used is partially transformed into heat. This means that the outputs are e-waste, which falls out of the system without being recycled and heat. In addition to these inputs and outputs, couplings between ICT and other economic sectors exist. In the following, we describe the main properties of the user devices and their internet connection, the user behavior, resource sources and waste sinks, and the coupling of ICT and other sectors from a resource use perspective. Afterwards, we analyze the system dynamics of ICT.

\subsection{Smartphones, Laptops, and Their Internet Connection}\label{sec:IctAsComplexSystem:SmartphonesInternet}

While the development and utilization of user devices, data centers and communication networks, and the behavior of the users have partially individual dynamics, also usage patterns exist that are relevant for multiple of these components. For example, most apps that users run on their smartphone or laptop employ a backhaul connection via the internet to servers, such that their usage affect all of these components.

Data centers and communication networks are built in the expectation of new apps or features occurring in the near future \cite{bloombergNEF}. However, the new apps and features can only be deployed once the according data centers or communication networks have become available. If new capabilities are added to communication networks and data centers, these are then employed within mobile apps, websites, and other applications, such as to enhance convenience and extend usage periods, which also leads to a higher energy consumption. As an example, the resolution of video streams has continuously increased over the last decades. Furthermore, real-time notifications, endless scrolling and gamification elements are used currently, which enhance flow and habit related to the apps, but also lead to mobile app addiction \cite{Jo2023,Seo2019}. The boundaries for this development depend on the social context, laws, the capabilities of communication networks and data centers, and the technological possibilities at the specific point of time. 

As an example, in wireless links, such as in 4G and 5G, multiple frequency bands are used for the data exchange. Within each of these frequency bands, regulatory bodies have implemented exposure constraints, which also limit the power of the data transmissions. Over the different cellular network generations, new frequency bands have been included into the wireless links. Out of the frequency bands used in previous network generations, most bands have continued being used for legacy reasons for a while, and later been added to the new mobile network generations. As the available data throughput increases with the deployment of new frequency bands, new applications become enabled which then leads to an increased demand for data rate \cite{WILLIAMS2022112033}. Hence, the combined use of the above frequency bands leads to an increase in energy use. Consequently, the deployment of new frequency bands and network equipment is identified to be the main driver of the increased energy use of mobile networks in \cite{EricssonEnergyCurve}. Accordingly, a study investigating communication networks in Finland found even at the very large throughput rates in Finland an increasing overall energy use despite the efficiency gains \cite{ICT4S2018:Energy_consumption_of_mobile}. We believe that the main mechanisms occurring here show similarities to those in transportation engineering, where the addition of lanes to roads lead to an increased car usage rather than reduced traffic jams (see \cite{MARCHETTI199475,gehl2013cities}).

Within data centers, the back-end software of applications is hosted. This includes data and files to be stored, processed, and transmitted to the user devices. Thereby, the data processing involves the training and application of neural networks. In \cite{Loten2023}, it is estimated that nowadays 20\% of the data center capacity is employed for artificial intelligence (AI). Within the corresponding data centers, the AI applications lead to an increased power usage \cite{StramDataCenterAIMarket}. The exact power usage depends on the architecture of the involved algorithms, the amount of data processed, and the employed hardware. While some tasks executed on data centers are directly related to the queries of users, other tasks such as backup generation and parts of the neural network training are executed periodically (such as daily or hourly) on the data centers \cite{bloombergNEF}.

When focusing on the current usage of smartphones and other user devices, the list of the most frequently downloaded apps on smartphones is dominated by social media apps and messaging apps \cite{statistaMobileAppsByDownloads}. Thereby, many of the social media apps contain the feature of endless scrolling through videos. Similarly, among the most frequently used websites, a high portion contains video content \cite{TopWebsiteRanking}. Consequently, video content was responsible for 75\% of data traffic in the internet in 2017, which was estimated by CISCO to increase to 82\% in 2022 \cite{CiscoVisualNetworkingIndex}\footnote{While the data cited here have originally been published in 2018, to the best of our knowledge no more recent data are available on the overall internet data traffic.}. Thereby, live internet video was predicted to be responsible for 17\% of the video traffic in 2022. According to the same report, internet gaming was expected to grow from 1\% to 4\%. Similarly, Ericsson estimates that video content has been responsible for 73\% of the traffic over cellular connections in 2023 \cite{EricssonMobilityReport2024}, and expects this value to increase when software and devices for augmented and virtual reality are further rolled out~\cite{EricssonMobilityReport}.

\subsection{The User Behavior and its Dependencies}

While ICT is used in a business context mainly to increase productivity, users employing ICT for personal use aim to obtain use value \cite{Marx1887}, e.g., through communicating with friends, gaining knowledge or being entertained. Hence, the ICT usage changes over the day \cite{10511857}. The specific behavior of a user depends on the technical capabilities of the available hardware, the economic situation of the users or companies, and the social context.

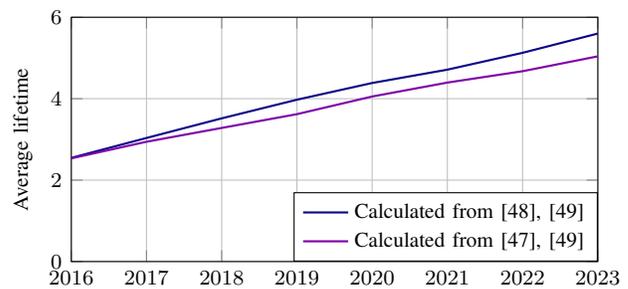
\begin{figure}%
    \centering
    \definecolor{mycolor1}{rgb}{0.97223,0.58172,0.25408}%
\definecolor{mycolor2}{rgb}{0.79638,0.27798,0.47132}%
\definecolor{mycolor3}{rgb}{0.49343,0.01152,0.65803}%
\definecolor{mycolor4}{rgb}{0.05038,0.02980,0.52797}%
\begin{tikzpicture}

\begin{axis}[%
    width=7cm,
    height=3.25cm,
    scale only axis,
xmin=2016,
xmax=2023,
ymin=0,
ymax=6,
xmajorgrids,
ymajorgrids,
    ylabel={Average lifetime},
legend style={at={(1,0)}, anchor=south east, legend cell align=left, align=left},
    x tick label style={/pgf/number format/.cd,%
          set thousands separator={},
          fixed},%
    font=\footnotesize,
]

\addplot [color=mycolor4,thick]
  table[row sep=crcr]{%
2016	2.54630322924907\\
2017	3.03543172075929\\
2018	3.51786013534719\\
2019	3.97536189339387\\
2020	4.38571006916788\\
2021	4.71325971102619\\
2022	5.12701405297563\\
2023	5.60045335767042\\
};
\addlegendentry{Calculated from \cite{statistaSmartphoneShipments,statistaSmartphoneSubscriptionsWorldwide}}

\addplot [color=mycolor3,thick]
  table[row sep=crcr]{%
2016	2.53437087749575\\
2017	2.94247985494933\\
2018	3.28211852421819\\
2019	3.62026736313981\\
2020	4.05393892161285\\
2021	4.39608482102864\\
2022	4.67504590261290\\
2023	5.04091102127990\\
};
\addlegendentry{Calculated from \cite{statistaSmartphoneSalesEndUsers,statistaSmartphoneSubscriptionsWorldwide}}

\end{axis}
\end{tikzpicture}%
    \caption{Average lifetime of smartphones over time, worldwide.}
    \label{fig:smartphones_useful_life_global}
\end{figure}

\begin{figure}%
    \centering
    \definecolor{mycolor1}{rgb}{0.97223,0.58172,0.25408}%
\definecolor{mycolor2}{rgb}{0.79638,0.27798,0.47132}%
\definecolor{mycolor3}{rgb}{0.49343,0.01152,0.65803}%
\definecolor{mycolor4}{rgb}{0.05038,0.02980,0.52797}%
\begin{tikzpicture}
\begin{axis}  
[  
    width=7cm,
    height=3.25cm,
    scale only axis,
    ybar,  
    ymin=0,
    ymax=8.5,
    ylabel={Average lifetime}, % the ylabel must precede a # symbol.  
    symbolic x coords={North America, Western Europe, China, Latin America, Sub-Saharan Africa, Middle East and North Africa}, % these are the specification of coordinates on the x-axis.  
    xtick=data,  
    ymajorgrids,
    xticklabel style={rotate=90,text width=1.5cm,align=right}, %Achsenbeschriftung drehen
    nodes near coords, % this command is used to mention the y-axis points on the top of the particular bar.  
    every node near coord/.append style={
        /pgf/number format/.cd,
        fixed,
        fixed zerofill,
        precision=2
    },
    %nodes near coords align={vertical}, 
    font=\footnotesize,
    bar width=0.75cm,
    xtick align=inside,
    ]  
\addplot[mycolor4,fill=mycolor4] coordinates {(North America,2.37896566205131) (Western Europe,3.35290743992171) (China,4.07090437244651) (Latin America,4.36635780828568) (Sub-Saharan Africa,4.56736200134138) (Middle East and North Africa,7.32436153570012) };  
%  2.31388888888889	3.07909090909091	4.22877192982456	4.40800000000000	7.07787500000000
\end{axis} 
\end{tikzpicture}
    \caption{Average lifetime of smartphones in different regions, 2022.}
    \label{fig:smartphones_useful_life_regional}
\end{figure}
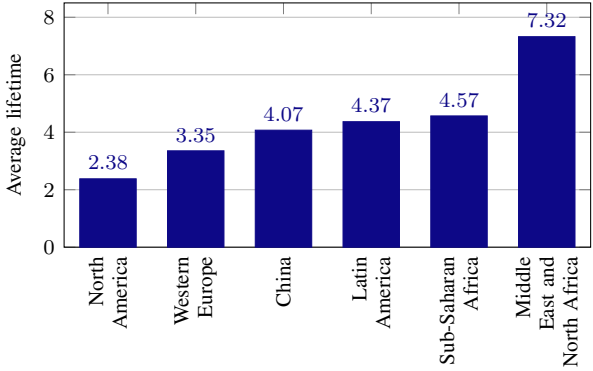

For instance, the lifetime of smartphones has significantly changed over time and varies across different regions. Based on the data for the number of devices sold worldwide from \cite{statistaSmartphoneSalesEndUsers,statistaSmartphoneShipments} and the number of mobile network subscriptions from \cite{statistaSmartphoneSubscriptionsWorldwide}, we can calculate how the average lifetime of smartphones has changed globally over time. Therefore, we assume that the devices in use are the ones most recently sold to calculate the average age of the devices used in a specific year. Based on this, we calculate the lifetime of the devices as twice the average age of the used devices. The results are shown in \figurename~\ref{fig:smartphones_useful_life_global} and indicate that the global average lifetime of smartphones has approximately doubled over the last few years.
Similarly, we calculate the lifetime in different regions of the world based on data from \cite{statistaSmartphoneSalesByRegion} and \cite{EricssonMobilityReport2024} and show the results in \figurename~\ref{fig:smartphones_useful_life_regional}\footnote{For years without available data on the number of smartphone sales, we assume the number of sales to equal those of the closest year with available data. Moreover, within the data sources, regions exist with different labels and thus limited comparability, which are mostly omitted. The number of smartphone sales in China is estimated from the value for Greater China through scaling by population.}.
Note that the numbers shown here do not cover that phones sold in one region might be resold in another region after their first use \cite{EuropeanEconomicAndSocialCommittee.CEPS.2019} and should be seen as approximations. Still, the figure shows a long average lifetime of smartphones in Africa and Latin America, and a short average lifetime in Northern America and Western Europe. These values have a negative correlation with the gross domestic product (GDP) per capita, and thus indicate a dependency between the smartphone use and the economic situation. Hence, we conclude that the increase of the average lifetime in \figurename~\ref{fig:smartphones_useful_life_global} comes at least partially from smartphones having become affordable in countries with a lower GDP, where a frequent replacement might not be affordable yet.

\begin{figure}
    \centering
    \input{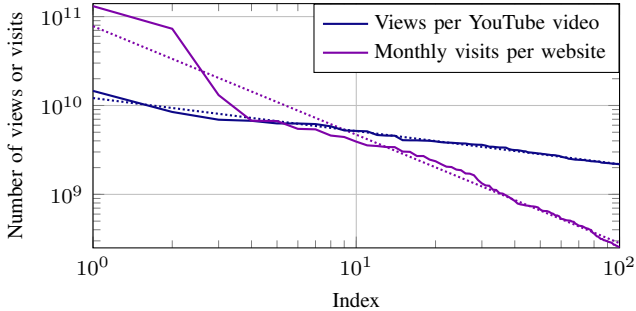}
    \caption{The number of views of the most-often viewed YouTube videos and the number of monthly visits of the largest websites for the 100 most viewed regarding visited videos or websites.}
    \label{fig:users_and_views_social_media_and_videos}
\end{figure}

Furthermore, users are likely influenced by their friends or other users. For instance, users might be interested in using the same messaging or social media platform that is also used by their friends to be able to mutually communicate. Hence, if a specific platform is used by more users, this is likely to attract additional users. Due to this effect, regional differences in social media use have significantly decreased over time \cite{WorldMapofSocialNetworks}, even if some regional differences still exist in the most famous messaging and social media platforms in each country \cite{MostPopularMessagingAppsWorldwide2023,WorldMapofSocialNetworks}. Similarly, the interaction of a user with a certain content or website often leads to the content being ranked higher by the social media app or search engine, such that the content is shown to additional users. Hence, the usage concentrates among a few websites and social media platforms, and some internet content ``goes viral'' \cite{Fieguth2021}, while other websites, platforms and content attract only a few users or remain completely unnoticed. For websites and platforms, this effect further strengthens as larger platforms can often gain larger profit, which can then be re-invested into the platform, and have more data available for neural network training. In \figurename~\ref{fig:users_and_views_social_media_and_videos}, we show for the most popular YouTube videos \cite{youtubeMostViews} and websites \cite{mostVisitedWebsites} the number of views and monthly visits, respectively, over the index of the video or website, sorted from the largest to the smallest. 
The straight line that the curve of the number of views per YouTube video follows in the log-log domain show that the power law distribution is an accurate model for this curve (see the discussion in \cite{Fieguth2021}). The curve of the monthly visits per website shows some deviations from the best fit straight lines at low indices\footnote{Within the data shown here, uncertainties might occur as famous websites are partially used via apps rather than the websites themselves.}, but the power-law distribution (or another heavy-tailed distribution) can still be used as an approximate. Such distributions are typical for complex systems \cite{Fieguth2021} and occur here due to the described self-reinforcing effect of the usage.

\subsection{Resource Sources and Waste Sinks}

During the manufacturing and operation of all ICT devices, significant amounts of electricity are used, which partially come from fossil fuels and thus leads to CO\textsubscript{2} and other greenhouse gas emissions. Thereby, the momentary energy consumption consists of a static component that determines the energy use at idle times and a dynamic component that depends on the workload of the devices \cite{09177e7695634f7ebf9e9b872cd7941a}. In addition, the device manufacturing requires various metals and other materials. For instance, within the manufacturing process of a smartphone, more than 50 different materials are used \cite{Fairphone2024}. After the end of the lifetime of the devices, only 22\% of small ICT equipment such as smartphones and personal computers and 25\% of screens are recycled \cite{GlobalEWasteMonitor2024}. The remaining part of the devices is collected outside of formal system or disposed in landfills, which has impacts on the corresponding environments and human health \cite{GlobalEWasteMonitor2024}. Note that these numbers do not include \emph{hibernating} devices, i.e., devices that people keep at home after the end of their usage. Due to the low recycling rates, the mining of new resources is required in the manufacturing process of new ICT devices. The ecological impacts from the resource mining, device manufacturing and e-waste are unequally distributed around the globe, affecting countries with a lower GDP more than countries with a higher GDP \cite{Lennerfors2015,GlobalEWasteMonitor2024}. Parts of the electricity used is transformed into heat, which leaves the system during the manufacturing and operation of all ICT devices. Within data centers, large amounts of the energy usage and heat emissions occur at concentrated locations. As AI data centers run at higher intensity than traditional data centers, these data centers require more energy than traditional data centers and water cooling is used \cite{StramDataCenterAIMarket}. The amount of energy and other resources that ICT requests depends on various aspects, of which we have already seen some in the previous subsections.

In various countries, recycling centers exists that take back old ICT devices and recycle various materials out of the devices. Furthermore, stores that sell smartphones and laptops often accept the return of devices at the end of the use phase. However, the usage of these recycling options depends on their convenience in comparison to the convenience of other options. Hence, larger ICT devices and data centers \cite{statistaDataCenterRecycling} are reused and recycled more frequently than smaller devices such as smartphones and laptops \cite{eurostatDestinationOfIctDevicesNoLongerInUse}. According to Eurostat \cite{eurostatDestinationOfIctDevicesNoLongerInUse}, only 10.35\% of EU27 citizens have recycled their old phone, while 9.86\% of people have recycled their old laptop and 12.8\% have recycled their desktop computer. Most of the people, i.e., 49.06\%, 32.55\% and 18.64\%, have kept their old phone, laptop or desktop, respectively, at home. 16.67\%, 11.31\%, and 8.47\% have sold or given away their previous phone, laptop, or desktop computer, respectively. 2.13\%, 1.44\%, and 2.10\%, respectively have disposed their previous devices into general waste, from which they cannot be recycled. Due to the high number of people who kept their old phones, 5-10 billion phones are hibernating in households worldwide after their end of use \cite{GSMA2023}. Hibernating devices that are recycled very late require a high amount of materials to be kept in the system. Thus, both hibernating devices and devices that end up in general waste both lead to an increased demand for new materials.

\subsection{Couplings of ICT and Other Sectors}

As the ICT capabilities progress, ICT transforms an increasing number of other economic sectors, such that a growing number of couplings between ICT and other sectors can be seen. For instance, the increased accessibility of ICT devices has lead to changes of text article reading and a reduced paper usage \cite{su10093027}. These changes are responsible for increasing carbon emissions in the ICT sector, and for decreasing carbon emissions at paper media. Similarly, the broad availability of streaming services has led to changes within the entertainment and media sector \cite{su10093027}. The internet of things consists of billions of small devices \cite{statistaNumberofInternetofThings}, which monitor their environment or perform specific actions and couple part production and ICT. Thereby, an enhanced efficiency in the production can be achieved, such as due to reducing downtime or optimizing the energy consumption in real time \cite{SOORI2023192}. This principle is referred to as an \emph{enabling effect} of ICT \cite{Santarius2023}, as ICT enables more sustainable practices in other sectors. Additional enabling effects exist in the fields of power supply, where smart meters can help reducing or shifting the energy use of households \cite{6520030}. The data communication of smart meters with the grid is typically implemented via power line communication (PLC) or, similar to other ICT devices, via radio frequency technologies. Moreover, with the uprising of connected cars, vehicles are increasingly connected to the internet, offering new opportunities for maintenance and the display of transportation-relevant information.

Within finance, ICT has made faster tradings on stock markets possible. Furthermore, payment systems are becoming increasingly digital, involving nowadays smart cards, online payments, and mobile applications \cite{Teker2022}. In the recent years, multiple cryptocurrencies have been developed, which have been suggested as a new form of currency. Some of these cryptocurrencies are implemented through computationally inefficient proof-of-work algorithms \cite{WENDL2023116530}. Due to these kinds of algorithms, special application-specific integrated circuits (ASICs) are required \cite{WENDL2023116530} and Bitcoin alone is estimated to have consumed in 2023 approximately \SI{121.13}{TWh} of electricity \cite{cambridgeBitcoin}, which is equivalent to \SI{0.47}{\%} of the global electricity consumption. Within these sectors, structural aspects are relevant going beyond those described here, which determine how the application of ICT impacts these sectors. 

\subsection{System Dynamics of ICT}

A system is complex, if the identities of the sub-systems and the whole are bi-directionally non-separable \cite{Estrada2023}. Especially, Morinian interactions \cite[Definition 3]{Estrada2023} between different sub-systems need to occur, i.e., interactions through which the nature of the different sub-systems changes. So far, we have seen that the capabilities of data centers, communication networks and user devices influence developments within apps, which further lead to changes in the user behavior. In turn, user behavior changes indicate potential for additional user behavior changes and, thus, impact the development of data centers and communication networks. Moreover, the user behavior influences the ranking of websites and other content in search engines and social media platforms, and, thus, has effects on the development of websites and apps. This means that various Morinian interactions between the different sub-systems of ICT are present. Furthermore, the definition of a complex system from \cite[Definition 4]{Estrada2023} is fulfilled.

Parts of the usage of ICT are bounded by the population on earth. Hence, the number of annual smartphone sales has already converged \cite{statistaSmartphoneSalesEndUsers,statistaSmartphoneShipments}, and the growth in the number of mobile network contracts has significantly declined in the recent years \cite{statistaSmartphoneSubscriptionsWorldwide}. However, due to a positive feedback loop between the deployment of new technology and changes of the user behavior, still a significant growth can be observed within other parts of the ICT system. For instance, the usage of video streaming services \cite{statistaVideoStraming} is continuously increasing, similar to the screen resolutions \cite{W3SchoolsScreenResolution} and the number of connected IoT devices \cite{statistaNumberofInternetofThings}.  Moreover, an increasing number of frequency bands is employed by cellular networks. As all real-world systems are stable, all growing dynamics will either converge or lead to collapse at a certain point \cite{Fieguth2021}. Here, the general limits to growth from \cite{Meadows1972} apply, covering the inputs of energy and non-energy resources, the availability of human labor, the size of industry capital that can be maintained, and different forms of waste/pollution. Furthermore, the discussion of Section~\ref{sec:IctAsComplexSystem:SmartphonesInternet} provides additional boundaries for the aforementioned dynamics, which emerge from physical limits of the employed technologies and various social aspects. Consequently, in Finland, where the overall mobile data traffic is relatively high \cite{ICT4S2018:Energy_consumption_of_mobile}, the growth rate of the traffic has decreased, representing now more a linear growth than an exponential growth \cite{trraficomVolumeCommunicationFinnland}. The same can also be observed in recent data of the worldwide mobile data traffic \cite{EricssonMobilityReport2024}. 

\section{Measures for Environmental Sustainability}\label{sec:MeasuresForEcologicSustainability}
The above system analysis can be employed to introduce measures for environmental sustainability covering the use phase and the production phase. For both phases, various efficiency, sufficiency, and consistency measures can be employed. In the following, we discuss various of such measures that aim to reduce the amount of non-renewable resources consumed.

\subsection{Efficiency Measures}\label{sec:efficiency}

The use phase energy is responsible for the largest proportion of the greenhouse gas emissions from data centers and communication networks \cite{BELKHIR2018448,MALMODIN2024102701}. Moreover, also relevant parts of the emissions from user devices occur during the use phase. To reduce the environmental impact in this phase, each ICT component can individually be designed more energy-efficient. This way, the energy usage is minimized under the assumption of a constant input. The studies investigating potential energy efficiency measures cover all different aspects of ICT. For example, energy efficiency techniques for 5G radio access networks have been surveyed in \cite{9678321}. Similarly, a survey on software designs for energy-efficient cloud computing has been undertaken in \cite{Katal2022}. These efficiency gains mitigated an increased energy demand due to growth of the usage to significant parts in the past \cite{IEADataCentersTransmissionNetworks}. Especially, the EnergyStar label and the EU energy label have made aspects on energy efficiency of the designs transparent to consumers to change their product choices and impact design decisions by manufacturers \cite{FREITAG2021100340}. The EU energy label is part of the EU energy-related products (ERP) directive, which shows that such labels can also form the basis for regulations of technology in a further step. However, with the slow-down of Moore's law, \cite{Andrae2020} predicts that the gains achieved due to efficiency measures are decreasing.

Nevertheless, it is still possible to achieve efficiency gains. Therefore, one option is to optimize the efficiency of multiple components jointly in addition to the individual optimization of each individual technical component \cite{ying2024usingwastefactoroptimize,9771701}. Moreover, a high focus is currently set on maximizing the efficiency of the devices at idle times (instead of just the peak times) \cite{10511857}. Furthermore, additional parameters such as the temperature can be taken into account to adapt technology configurations in the favor of efficiency \cite{09177e7695634f7ebf9e9b872cd7941a}. For wireless communication links, there is a wide discussion on the employment of optimized beamforming strategies such as non-orthogonal multiple access coding and rate-splitting \cite{10323352,9813735}. Moreover, discussions have emerged on the modification of the wireless channels such that transmissions need less energy, such as via intelligent reflective surfaces \cite{10348506,10323352,9759225}. However, where such techniques require changes of standards, it is expected that a longer time is needed until the gains of such measures are obtained. Also, further improvements of the sustainability labels are discussed, such as via labels for the efficiency of software \cite{Kern2015/09}.

Among the structural aspects analyzed before, efficiency measures can take the distribution of the data requested by the user and the time at which users are active into account. Since HTTP/1.1 \cite{http-rfc}, frequently accessed files can be cached directly at the user's web browser, which reduces the data transmissions required for these files and thus leads to a more efficient utilization of the communication medium. As different users in the same region demand access to the same content, nowadays edge data centers are being built. For instance, in the internet of things, data are monitored and accessed only from within the same region, such that a computing in the edge data center can reduce wide-range transmissions. Furthermore, data can also be cached in regional data centers, which is, e.g., done within Google's project amplified mobile pages (AMP) to reduce latency. Such approaches lead to additional energy and non-energy resources being required for the operation of the regional data centers, but reduce the energy required for data transmission. Overall, \cite{8668812} found that distributed architectures for fog and edge computing can indeed lead to an increased energy efficiency for some applications. 

\subsection{Sufficiency Measures}

Besides the efficiency, the resource consumption of ICT depends on the intensity of the usage. Hence, sufficiency measures can be applied within the design of apps and websites with their back-end software to avoid resource-intensive user behaviors and help achieving a low energy consumption. Thereby, due to the distribution of the accessed content shown in \figurename~\ref{fig:users_and_views_social_media_and_videos}, it is most crucial to focus on frequently accessed platforms and content, while measures for less-frequently accessed websites still have an effect. In this context, the EU commission has reached an agreement with Netflix, Inc., at the beginning of the Covid-19 pandemic to reduce network congestion from video streams through reduced video resolution \cite{Sweney2020}. Similar measured can also employed for environmental sustainability. As most users do not change the standard video resolution, one option is to select lower streaming resolutions as default settings \cite{10.1145/3401335.3401709}. Thereby, as the the energy consumption depends on the screen size \cite{su10093027}, it should be ensured that streaming from smartphones and laptops remains more convenient than watching TV. Moreover, a reduced duration of the video content consumption can be targeted, which, according to \cite{IEAStreaming}, can even have a larger impact on the energy usage than a reduced video resolution. To tackle the problem of mobile app addiction, the operating systems Android and iOS already include software for digital well-being and screen time management. This software should be improved and made more easily accessible. Especially, an improved data reporting can be used to further enhance these programs. For instance, if a high share of users activates a digital well-being software for a specific app, the operating system might automatically suggest other users with high usage times to enable such a functionality\footnote{It might here be important to consider relative and absolute numbers to avoid triggering this feature for very small apps with a few dozen users.}. Furthermore, software features that often lead to mobile app addiction might be regulated for large platforms.

Additional sufficiency measures can be introduced within the design of data centers and communication networks, which aim to reduce the energy consumption across the full chain of ICT devices. Data centers are often built in or nearby big cities, where the available area is limited. Hence, various cities have undertaken measures to limit the construction of new data centers \cite{bloombergNEF}. Within the communication networks, a two-sided pricing system dependent on the network usage already provides up to a certain extend incentive to app developers to utilize communication networks efficiently and sufficiently \cite{Chiang_2012}. For the mobile links near the users, we have seen in Section~\ref{sec:IctAsComplexSystem} that the continuous addition of new frequency bands to cellular networks leads to an increasing energy demand, but can also contribute towards enabling effects. Here, an option could be to replace volume-limited mobile network contracts by rate-limited mobile network contracts. This would lead to an automatic reduction of the resolution of video streams, while having only a small impact on text-based content and enabling effects. Note that, while the capabilities of base stations and data centers are typically optimized for peak-times, i.e., times with high utilization, the utilization is only moderate or low at larger parts of the time \cite{10511857}. Hence, by encouraging users to shift their usage to off-peak times, the amount of hardware being required at the base stations and data centers can be reduced. To do so, mobile phone contracts have introduced off-peak and time-dependent contracts within 3G mobile networks \cite{EricssonMobilityReport,Chiang_2012}. With the roll-out of 5G, these contracts later became unavailable as the new 5G systems automatically distribute the maximum data rate among the users.

In addition to the operation, also the manufacturing of the ICT devices is responsible for a significant share of the greenhouse gas emissions. Especially, this share is large for the user devices  \cite{FREITAG2021100340,BELKHIR2018448,MALMODIN2024102701}. Hence, enhancing the time over which these devices are used can significantly contribute to reducing the average annual emissions and material usage. Due to the regional differences shown in \figurename~\ref{fig:smartphones_useful_life_regional}, this is especially relevant in countries where a frequent replacement of the devices is affordable. According to the works \cite{CORDELLA2021125388,EuropeanEconomicAndSocialCommittee.CEPS.2019,WILSON2017521}, the main reasons for phones being replaced that should be addressed are as follows:
\begin{itemize}
    \item The available phones are often not fully functioning any more, such as due to battery failures or a broken screen, and repairs are not available. 
    \item Only low knowledge or information on how and where phones can be repaired are available.
    \item The use of repair services is partially too complicated or expensive and considered to be not worth the effort.
    \item The availability of software updates is often limited to short time-frames, and people want to have the latest software.
    \item Some mobile network contracts include a regular supply with new phones, which also leads to reduced usage periods. Moreover, the conclusion of a new mobile network contracts can also lead to people buying new phones. 
    \item In some cases, people also want to have a new phone, due to new features, a new smartphone design, or because the previous phone looks old. This behavior might be influenced by social pressure or advertisements that target a frequent replacement rather than long use or repairs.
    \item The previous phone has been lost or stolen.
\end{itemize}
This means that often the lifetime of the devices can be enhanced by increasing the durability of the hardware design in terms of reliability and/or repairability. Note that all components of a phone are typically engineered to last over a similar usage duration before becoming defect, which is determined by the warranty time. Hence, extensions of the legal warranty time requirement would also lead to a more robust design of various components of a phone. One option to achieve durability is through a modular phone design. For instance, due to a modular design optimized for longevity, phones from Fairphone can directly be repaired by the user in many cases, such that these phones have a longer lifetime than average phones in the same markets \cite{Fairphone2024}. Note that while a long-lasting phone design is likely to increase the cost of each single purchase of a phone, users can save money if they need to replace their phone less often due to such a design.

To adapt the sustainability measures to future trends in ICT, a continuous monitoring of new resource-intensive applications and their impacts on the user behavior would be helpful. In addition to sufficiency measures targeting explicitly ICT, also economy-wide sufficiency measures can contribute to reducing ICT's environmental implications. A general analysis of such measures is outside of the scope of this paper. Still, we would like to highlight that carbon taxes and emission trading systems (ETSs) active in many countries can help here \cite{sru2024,Santarius2022}. Note that for these measures, a socially just implementation is crucial that ensures a sufficient technology access for everyone.

\subsection{Additional Consistency Measures} 

As a reduced resource consumption simplifies the transition to closed material circles and renewable energy, the aforementioned efficiency and sufficiency measures can also be part of consistency measures. Moreover, by taking the full resource flows into account, additional consistency measures can be formulated.

The energy used comes from fossil, nuclear, or renewable sources and leaves the system as heat. To change the energy supply to renewable energy, ICT companies can install their own renewable electricity generation systems or purchase renewable energy from a third-party contractor \cite{HUANG2020114109}. Thereby, a transition to a 24/7 coverage of the renewable energy production and consumption should be targeted \cite{IEADataCentersTransmissionNetworks}. However, in the case of electricity from solar panels or wind turbines, the electricity generation is time-dependent. Hence, adaptations of the point of time at which electricity is consumed can help in this transition by reducing the need for battery storage. In this context, ICT-based solutions have been developed that make the energy (or other resource) consumption of households visible to users along with the energy production of locally installed solar panels. By employing such solutions, users can be encouraged to shift the usage of energy-intense devices such as washing machines and dishwashers to times with more available renewable energy~\cite{5768e477ee1843d5bdf68d4b508cbecb}. Heat pumps and electric cars can nowadays even automatically adapt their demand depending on local electricity generation or momentous electricity prices.

Similarly, also ICT devices themselves can be coupled to the availability of renewable energy to provide a flexible energy demand. This holds especially for data centers, which can provide flexibility by shifting the execution of specific tasks in time and location \cite{bloombergNEF}. For example, the generation of backups and the training of AI applications can often be shifted by a few hours to times at which more renewable energy is available in the grid or generated locally. The scheduling of such tasks can first be optimized via the day-ahead electricity prices, the predicted local electricity generation and the predicted demand, and then be adapted in real time to deviations from the predictions \cite{9078781}. Moreover, companies operating multiple hyperscale data centers sometimes store frequently accessed data at multiple locations or have copies of neural networks. In this case, the according companies might be able to shift tasks among data centers depending on the momentary energy costs at both data centers \cite{CIOARA2019392,bloombergNEF}. Moreover, also the edge data centers built currently may contribute to flexible demand~\cite{10.1145/3484266.3487394}.

The waste heat, which leaves the system at the different ICT components, has traditionally not been further used. Within data centers, a large amount of heat leaves the system at a single point. Collecting and using this heat, such as within nearby buildings or by feeding it into district heat systems, can help decarbonizing the heat supply \cite{Romero2014/08}. In principal, waste heat from the data centers can be collected in the form of hot water for all kinds of cooling systems currently in use. However, the different kinds of cooling systems have different quality for a further usage in heating system. The two-phase cooling has the highest quality, followed by water cooling and air cooling \cite{HUANG2020114109}. Still, in most cases, the temperature of the water collected from the servers is too low to be fed directly into the district heat networks. Instead, heat pumps can be used to leverage the water temperature by compressing the heat, and eventually by further increasing the water temperature through geothermal energy sources to be in line with the district heating requirements \cite{HUANG2020114109}.

Within the production, the greenhouse gas emissions and other environmental impacts depend on the amount of new natural resources needed for each produced device. Keeping the non-energy resources that are part of the different ICT components inside of the system can decrease the amount of new natural resources used by ICT and the amount of e-waste produced. Therefore, it is crucial to avoid that devices end up in general waste or landfills and to increase recycling rates. Furthermore, the duration over which phones are hibernating should be minimized. Therefore, various reasons should be addressed that lead to devices not being recycled (see e.g., \cite{WILSON2017521,EuropeanEconomicAndSocialCommittee.CEPS.2019,BAI2018227}):
\begin{itemize}
    \item Recycling points are often complicated to use, such as due to being far away from the places people visit for daily routines.
    \item People are not aware of existing recycling options \cite{EuropeanEconomicAndSocialCommittee.CEPS.2019}.
    \item Some people have concerns regarding possible personal information disclosure \cite{BAI2018227}.
    \item If devices are still functioning to a certain extent, they are often kept as backup device, such as for occasions where a phone gets more likely stolen or for the case that the primary phone gets broken \cite{WILSON2017521}. However, \cite{WILSON2017521} points out that likely only one old phone is used as backup phone.
    \item Devices sometimes still have a sentimental value for people \cite{WILSON2017521}.
    \item Benefits of recycling devices, such as environmental or financial, are considered low. Moreover, especially smartphones only take limited space at peoples home, such that freeing this space is partially not considered relevant.
\end{itemize}

To increase recycling rates, easily accessible recycling options have been suggested, such as strategically placed recycling options \cite{EuropeanEconomicAndSocialCommittee.CEPS.2019}. For instance, recycling bins for electronic devices could be placed at locations frequently attended by a large group of people, such as supermarkets or stations. Also, device-as-a-service contracts, i.e., contracts where the devices are lent to the users, can increase recycling rates \cite{Santarius2022,Fairphone2024}. Within such, the devices remain owned by a network operator or smartphone manufacturer, which also handles the recycling after the end of the usage. Moreover, financial benefits can be provided for the return of devices \cite{EuropeanEconomicAndSocialCommittee.CEPS.2019}. For instance, some companies offer rewards when returning an old smartphone at the purchase of a new one \cite{BAI2018227}. Furthermore, some companies have created business models around easily accessible recycling services for smartphones and advertisement campaigns for these recycling services \cite{BAI2018227}. In all of these approaches, we believe that besides showing the environmental value of recycling, it is important that transparency is provided on the recycling process.

\section{Limitations of this Study}

Within the scope of this work, we have focused on environmentally sustainable ICT in terms of SDGs 12, 13 and 15. Furthermore, we have focused on the behavior of the users related to the direct environmental impacts of the ICT devices. However, we have only marginally touched on social sustainability and the application of ICT as enabler for environmental sustainability in other areas. For instance, if, e.g., the lifetime of phones is extended, it is likely that less workers are needed in the manufacturing process. Within our work, no analyses on the impacts on employment have been undertaken. 
Moreover, we have not considered methods for the reduction of the environmental impact that are only indirectly related to the operation of the ICT devices, such as for maintenance, the operations of the buildings in which data centers are hosted or the shipping of devices.
Furthermore, it should be noted that parts of the above analysis and suggestions focus on qualitative (rather than quantitative) effects. 

\section{Discussion and Recommendations}

The ICT sector can contribute to the fulfillment of the goals set by the international community. Therefore, we have identified various social and technical aspects, which the user behavior and the environmental impact of ICT depend on. More specifically, the description of ICT as a complex system has shown that structural dynamics behind the usage patterns exist, whose understanding can help in achieving an environmentally sustainable ICT. Based on the structural analysis, we have discussed various sustainability measures, covering aspects of efficiency, sufficiency and consistency. For the implementation of such measures, action is required from policy makers and industry.

Due to the time-criticality of the sustainability targets, the implementation of sustainability measures should not be delayed until further research is available. Nevertheless, future research can be helpful to adapt the sustainability measures as research develops. Therefore, we propose some recommendations for future research in the following:
\begin{itemize}
    \item In Section~\ref{sec:MeasuresForEcologicSustainability}, we have described various measures for environmental sustainability. For most of the discussed measures, already some kinds of implementation or broader documentation exists. However, these implementations are often not optimized for the structural aspects of the ICT usage, and some of the discussed measures have originally been applied outside of the context of environmental sustainability. A utilization of the discussed structural aspects would allow further optimization of these technologies for sustainability. Also, a broader research on possible designs of sufficiency and consistency measures for the use phase could be useful. 
    \item Similar to parts of the related works, we have noticed a lack of up-to-date data describing usage patterns of ICT systems during production phase, use phase, and the devices' end of life. We believe that it would be relevant to fill the according gaps within the data to better understand usage patterns and their changes over time. For instance, data from operating systems designers, network operators, and surveys covering the user behavior could help here.
    \item In this work, we have described the ICT sector as a complex system. Further optimizations of the model description and a mathematical formalization of parts of the model can help in the design of environmentally sustainable ICT. Furthermore, as the corresponding technologies have a broad use within other sectors, a complex system analysis of these sectors would be helpful to determine how ICT can optimally assist the social-ecological transformation of these sectors.
\end{itemize}

\ifCLASSOPTIONcaptionsoff
  \newpage
\fi

\bibliographystyle{IEEEtran}
\bibliography{main}

\begin{IEEEbiography}[{\includegraphics[width=1in,height=1.25in,clip,keepaspectratio]{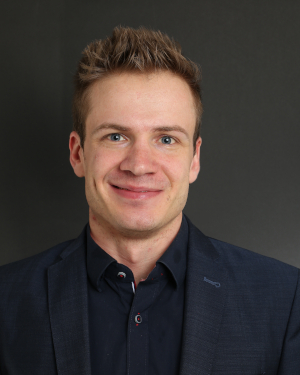}}]{Stefan Roth}
received the B.Sc., M.Sc., and Dr.-Ing. degree in Electrical Engineering and Information Technology from Ruhr University Bochum, Germany, in 2016, 2019 and 2024, respectively. Since 2019, he is working at the institute of digital communication systems at Ruhr University Bochum, where he is currently a postdoctoral researcher. His research includes the topics of sustainable ICT, physical layer security and communication in the internet of things.
\end{IEEEbiography}

\begin{IEEEbiography}[{\includegraphics[width=1in,height=1.25in,clip,keepaspectratio]{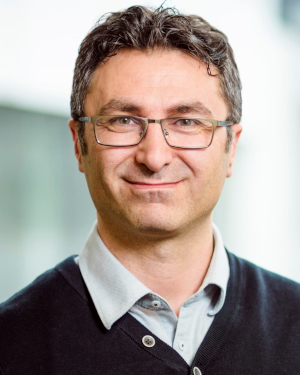}}]{Aydin Sezgin} (S'01--M'05--SM'13) received the Dr. Ing. (Ph.D.) degree in electrical engineering from TU Berlin, in 2005. From 2001 to 2006, he was with the Heinrich-Hertz-Institut, Berlin. From 2006 to 2008, he held a postdoctoral position, and was also a lecturer with the Information Systems Laboratory, Department of Electrical Engineering, Stanford University, Stanford, CA, USA. From 2008 to 2009, he held a postdoctoral position with the Department of Electrical Engineering and Computer Science, University of California, Irvine, CA, USA. From 2009 to 2011, he was the Head of the Emmy-Noether-Research Group on Wireless Networks, Ulm University. In 2011, he joined TU Darmstadt, Germany, as a professor. He is currently a professor with the Ruhr-Universit\"at Bochum, Germany. He has published several book chapters, more than 70 journals and 200 conference papers in these topics. Aydin is a winner of the ITG-Sponsorship Award, in 2006. He was a first recipient of the prestigious Emmy-Noether Grant by the German Research Foundation in communication engineering, in 2009. He has coauthored papers that received the Best Poster Award at the IEEE Communication Theory Workshop, in 2011, the Best Paper Award at ICCSPA, in 2015, at ICC, in 2019, and at ISAP, in 2023.

\end{IEEEbiography}

\end{document}